\documentclass[pra,aps,twocolumn,superscriptaddress,showpacs,tightenlines,amsmath]{revtex4}
\usepackage{epsfig}
\usepackage{amsmath,amssymb}

\begin{document}

\title{Entanglement and control operations in Ising interactions of bipartite qubits}

\author{Francisco Delgado}
\email{fdelgado@itesm.mx}
\affiliation{Mathematics and Physics Department, Quantum Information Processing Group, Tecnologico de Monterrey, Campus Estado de Mexico, Atizapan, Estado de Mexico, CP. 52926, Mexico.}

\date{\today}

\begin{abstract}

Entanglement generated by Ising model has been studied for several authors in order to understand the relation between it and magnetic properties of materials, principally using one or two dimensional models for two or more particles. In this work, Ising model evolution is solved in three dimensions for two parts including an inhomogeneous magnetic field, giving an detailed study of the entanglement properties derived from interaction. Some relations between entanglement and energy or spin are developed specifically to stablish candidates for entanglement witness. Finally, some basic quantum control operations are prescripted for this model as to preserve the properties of the system as for transfer information between the two parts. These last schemes of control are useful when initial state is completely known, to gain domain on the system. It prevents to introduce more sophisticated control schemes which normally are necessary when initial state is at least partially unknown.



\end{abstract} 

\maketitle

\section{Introduction}

Entanglement is used in quantum computation as a central aspect for improve information processing in order to exploding the interesting properties of quantum mechanics \cite{schrod1, schrod2}. For this reason, entanglement is subjected to deep research to understand completely its complexity, properties and potential usefulness \cite{bennet1, bennet2, nielsen1}. In the last sense, its control is one of the most important aspects of it \cite{branczyk1, xi1}, nevertheless its study will not have a complete map of road until his quantification and behavior could be understood. 

Nielsen \cite{nielsen2} was the first to report studies of entanglement in magnetic systems based on a two spin systems using the Ising model with an external magnetic field. After of this, different studies have extended this research  for more complex systems and depending on external parameters (as temperature and strength of external field) \cite{arnesen1, wang1}, and considering different models of Ising interaction (XX, XY, XYZ depending on focus given by each author in order to reproduce calculations related with one or two dimensional lattices) \cite{wang2, kamta1, sun1, zhou1, gunlycke1}.

In this paper we just study bipartite systems in three dimensions, but more in the aim of learn about entanglement and control related with Ising interaction and information transfer processes, than to study lattice properties of entanglement. Extensions to multipartite systems will be studied in future works.

\section{Ising interaction and variants}

Ising model is motivated mainly by far-field strength of a magnetic dipole interaction between two particles in which the energy of bounding is given by:

\begin{equation} E=\frac {\mu_0}{4 \pi r^3}(\vec{\mathbf{m}}_1 \cdot \vec{\mathbf{m}}_2-3\vec{\mathbf{m}}_1 \cdot \hat{\mathbf{r}} \vec{\mathbf{m}}_2 \cdot \hat{\mathbf{r}})  \end{equation} 

\noindent where $\vec{\mathbf{r}}$ is the vector distance between particles, $\hat{\mathbf{r}}$ its unitary vector associated and $\vec{\mathbf{m}}_i$ is the magnetic momentum of particle $\it{i}$. In addition, a model which relates the magnetic momentum with the spin is:

\begin{equation} \vec{\mathbf{m}}_i = \mathbf{g}_i \cdot \vec{\mathbf{s}}_i \end{equation}

\noindent with $ \mathbf{g}_i$ as a tensor. Combining two last expressions we obtain:

\begin{eqnarray} 
E&=&\frac {\mu_0}{4 \pi r^3} \sum_{j,l=1}^3(\sum_{i=1}^3 {g_1}_{ij} {g_2}_{il} 
-3 \sum_{i,k=1}^3 {g_1}_{ij} {g_2}_{kl} r_i r_k){s_1}_j {s_2}_l \nonumber \\ 
& \equiv & -J_{jl} {s_1}_j {s_2}_l  
\end{eqnarray}

Depending on ${g_k}_{ij}$ values the Ising-like interaction become in different interaction models which have been proved by several authors in studies of entanglement \cite{wang2, kamta1, sun1, zhou1, gunlycke1}. Ising model is precisely which where $J_{jl}$ is proportional to the identity (see appendix A):

\begin{equation} \label{ising}
E=-J \vec{\mathbf{s}}_1 \cdot \vec{\mathbf{s}}_2
\end{equation}

This kind of interactions were first used in statistical physics to describe the magnetic behavior of lattices in different ways precisely by Ising \cite{ising1, brush1} and after by Heisenberg \cite{baxter1} in quantum mechanics introducing Pauli matrices proportional to $\vec{ \bf r}_i$. Works listed before \cite{wang2, kamta1, sun1, zhou1, gunlycke1} and other more recent showing transference and control of entanglement in bipartite qubits \cite{terzis1} and lattices \cite{stelma1, novotny1} normally focus on describe magnetic properties of materials related with entanglement.  By other hand, Cai \cite{cai1} has considered a more general model with $J_{jl}$ diagonal in order to study the relation between entanglement and local information. Closer this last, but related with the previous papers, this work is focused on the full Ising model in three dimensions, in order to explore some properties of entanglement which arise between particles inclusively by adding an interaction with an inhomogeneous external magnetic field in the $z$ direction:

\begin{equation} \label{hamiltonian}
H= -J {\vec{\bf \sigma}}_1 \cdot {\vec{\bf \sigma}}_2+B_1 {\sigma_1}_z +B_2 {\sigma_2}_z
\end{equation} 

\noindent and some operations that one could apply to control entanglement.

Using explicit form of Pauli matrices we obtain in matrix form for (\ref{hamiltonian}):

\begin{equation} \label{mathamiltonian}
H= \left(
\begin{array}{cccc}
B_+-J & 0          & 0           & 0  \\
0         & B_-+J  & -2J         & 0 \\
0         & -2J        & -B_-+J  & 0  \\
0         & 0          & 0           & -B_+-J 
\end{array}
\right) 
\end{equation}

\noindent where $B_+=B_1+B_2$,  $B_-=B_1-B_2$, $R=\sqrt{B_-^2+4J^2}$. One more suitable selection of parameters is (to make finite some parameters and to reduce some expressions): 

\begin{eqnarray}
& b_+=B_+/R, b_-=B_-/R \in [-1,1], \nonumber \\  & j=J/R \in [0,1/2],  t'=Rt
\end{eqnarray}

We will not drop the prime in the time because the letters should be advertise to the reader about the actual selection of parameters. We use it in all cases which it is situable, except for some final expressions.

Diagonalizing the Hamiltonian (\ref{mathamiltonian}), we can obtain the operator of evolution, which become in Dirac notation:

\begin{eqnarray} \label{evolop}
U(t')&=&e^{-it'(b_+-j)} \left| 0_10_2 \right> \left< 0_10_2 \right| \nonumber \\ \nonumber
&&+ e^{-it'j}(\cos t'-i b_- \sin t') \left| 0_11_2 \right> \left< 0_11_2 \right| \\ \nonumber
&&+ i 2j e^{-it'j}\sin t' \left| 0_11_2 \right> \left< 1_10_2 \right| \\ \nonumber
&&+ i 2j e^{-it'j}\sin t' \left| 1_10_2 \right> \left< 0_11_2 \right| \\ \nonumber
&&+ e^{-it'j}(\cos t'+i b_- \sin t') \left| 1_10_2 \right> \left< 1_10_2 \right| \\ \nonumber
&&+e^{it'(b_++j)} \left| 1_11_2 \right> \left< 1_11_2 \right| \\ 
\end{eqnarray}

\section{Evolution and properties in the Ising model for 2-parts}
\subsection{Generalities}

It is well known that Ising-like interaction generate entanglement \cite{ wang2}. Our Hamiltonian has the eigenvalues (Fig. 1):

\begin{eqnarray} 
E_1&=&-J-B_+ \nonumber \\ \nonumber
E_2&=&-J+B_+ \\ \nonumber
E_3&=&J-R \\ 
E_4&=&J+R 
\end{eqnarray}

\noindent with the eigenvectors:

\begin{eqnarray} \label{eigenstates}
\left| u_1 \right> &=& \left| 0_1 0_2 \right> \nonumber \\ \nonumber
\left| u_2 \right> &=& \left| 1_1 1_2 \right> \\ \nonumber
\left| u_3 \right> &=& \sqrt{2} j \left( \frac{\left| 0_1 1_2 \right>}{\sqrt{1+b_-}} + \frac{\left| 1_1 0_2 \right> }{\sqrt {1-b_-}} \right)  \\ 
\left| u_4 \right> &=& \sqrt{2} j \left( \frac {\left| 0_1 1_2 \right>}{\sqrt{1-b_-}} -\frac{\left| 1_1 0_2 \right> }{\sqrt{1+b_-}}\right)
\end{eqnarray}

\noindent because of that, these states are invariant under interaction (\ref{evolop}). 

\subsection{Energy, spin and entanglement}

Spected values of energy and spin are normally related with entanglement.  Note first that just for some selection of parameters (when $b_- \to 0 \Rightarrow j \to 1/2$), the two last states of (\ref{eigenstates})  will correspond to maximally entangled states: $\left| u_3 \right>=\left| \beta_{01} \right>$ and $\left| u_4 \right>=\left| \beta_{10} \right>$, it means in an homogeneous field, this Bell states in $z$ direction basis are invariant. Present section present some remarks about relation between energy and spin with entanglement. We will begin since the most simple until most general scenarios.

Firstly, in absence of field one can show by direct calculation that for $\left| \psi \right>$ separable, then (see appendix B):

\begin{equation}
E=\left< \psi \right| H \left| \psi \right> \in [-J,J]
\end{equation}

\noindent it means that Hamiltonian could be used to define an entanglement witness operator because of $\left< \beta_{10} \left| H \right| \beta_{10} \right> = 3J$. Effectively, in this case 
$\left|0_1 0_2 \right>$, $\left|1_1 1_2 \right>$ (or alternatively, $\left|\beta_{00}\right>$, $\left|\beta_{11}\right>$), $\left|\beta_{01}\right>$, form an invariant subespace of  ${\mathcal H}^{\otimes 2}$ (all state obtained by combine this states is invariant) isolated from  $\left| \beta_{10}\right>$. In this case, energy is degenerated for any state of three first and it is $E_{1,2,3}=-J$, instead for the fourth state it's $E_4=3J$. In absence of field, non invariant evolution it's only possible by combining both subspaces.

After, by adding an homogeneous field, two of the three first degenerated eigenstates become unfolded with energies $E_1=-J-B_+$, $E_2=-J+B_+$ (depend on magnetic field, $B=B_+/2$) with separable eigenstates $\left|0_1 0_2 \right>$, $\left|1_1 1_2 \right>$ respectively. Instead $E_3=-J$, $E_4=3J$ remain unchanged and independent of magnetic field with the same entangled eigenstates $\left| \beta_{01} \right>$ and $\left| \beta_{10} \right>$, which are now the only invariant Bell states in $z$ direction basis. This apparent difference between influence or non influence on field appear as reason to get or not to get invariant separable eigenstates under evolution operator (\ref{evolop}). 

In the general case when magnetic field is inhomogeneous, energy of the two separable eigenstates unfolded, remain now unchanged. Nevertheless, the previous maximally entangled eigenstates become just partially entangled with energies $E_3=J-R \le -J$, $E_4=J+R \ge 3J$. Otherwise, one can prove than the two last eigenvectors, $\left|u_3 \right>$ and  $\left|u_4 \right>$, are not separable at least that  $|B_-| \to \infty$ \footnote{One can show that Schmidt coefficients of  $\left|u_3 \right>$,  $\left|u_4 \right>$ are $\lambda_{1,2}=\frac{1}{2} (1 \pm \frac{|B_-|}{R}) \to 1,0$ in this case. Additionally as $\lim_{B_- \to \infty} R(R \pm B_-)=\infty, 2J^2$ and $\lim_{B_- \to -\infty} R(R \pm B_-)= 2J^2, \infty$ then eigenvectors become: $\left| 0_1 1_2 \right>$, $\left|1_1 0_2 \right>$.}. By direct calculation one can show that for $\left| \psi \right>$ separable (see appendix B):

\begin{equation}
E=\left< \psi \right| H \left| \psi \right> \in [-M,M]
\end{equation}

\noindent where $M={\rm Max}\{J+|B_-|,|B_+| \}$. In this case, energy is no more in all cases an entanglement witness because for $\left< \beta_{1,0} \right| H \left| \beta_{1,0} \right>=3J+B_- \equiv E_{10}$. With this, one of $M$, $E_4$ and $E_{10}$ could be greater values than other, depending on values of $B_-$ and $B_+$. In some sense, complexity and difference in intensity of the magnetic field appear to destroy this relation energy-entanglement.

By comparison, the eigenvalues of ${\vec{\bf \sigma}}_1 \cdot {\vec{\bf \sigma}}_2$ are:

\begin{equation} 
\Lambda_{1,2,3}=1, \Lambda_4=-3
\end{equation}

\noindent with the eigenvectors:

\begin{eqnarray} \left| v_1 \right> &=& \left| 0_1 0_2 \right>  \nonumber \\ \nonumber
\left| v_2 \right> &=& \left| 1_1 1_2 \right> \\ \nonumber
\left| v_3 \right> &=& \frac{1}{\sqrt{2}} (\left| 0_1 1_2 \right> + \left| 1_1 0_2 \right> )= \left|\beta_{01} \right> \\ 
\left| v_4 \right> &=& \frac{1}{\sqrt{2}} (\left| 0_1 1_2 \right> - \left| 1_1 0_2 \right> )= \left|\beta_{10} \right> 
\end{eqnarray}

An important issue is that if some state $\left| \psi \right>$ is separable, then:

\begin{equation}
\left< {\vec{\bf \sigma}}_1 \cdot {\vec{\bf \sigma}}_2 \right> = \left< \psi \right| {\vec{\bf \sigma}}_1 \cdot {\vec{\bf \sigma}}_2 \left| \psi \right> \in [-1,1]
\end{equation}

\noindent so ${\vec{\bf \sigma}}_1 \cdot {\vec{\bf \sigma}}_2$ can be used to construct an entanglement witness operator in this case because (because of his proportional dependence with energy seeing before for abscence of field case):

\begin{equation}
\left< {\vec{\bf \sigma}}_1 \cdot {\vec{\bf \sigma}}_2 \right>=\left< \beta_{01} \right| {\vec{\bf \sigma}}_1 \cdot {\vec{\bf \sigma}}_2 \left| \beta_{01}\right>=-3
\end{equation}

The evolution (\ref{evolop}) of $\left< {\vec{\bf \sigma}}_1 \cdot {\vec{\bf \sigma}}_2 \right>$ become dependent just of $B_-$:

\begin{eqnarray} \label{evolspin}
\lefteqn{{\vec{\bf \sigma}}_1 \cdot {\vec{\bf \sigma}}_2 (t') = U^\dagger (t') {\vec{\bf \sigma}}_1 \cdot {\vec{\bf \sigma}}_2  U(t') =} \nonumber \\
&& \left| 0_1 0_2 \right> \left< 0_1 0_2 \right| + \nonumber \\
&& -(1+8jb_- \sin^2 t' )\left| 0_1 1_2 \right> \left< 0_1 1_2 \right| + \nonumber \\
&& 2(1-2b_-^2 \sin^2 t'+ib_- \sin 2t') \left| 0_1 1_2 \right> \left< 1_1 0_2 \right| + \nonumber \\
&& 2(1-2b_-^2 \sin^2 t'-ib_- \sin 2t') \left| 1_1 0_2 \right> \left< 0_1 1_2 \right| + \nonumber \\
&& -(1+8jb_- \sin^2 t' )\left| 1_1 0_2 \right> \left< 1_1 0_2 \right| + \nonumber \\
&& \left| 1_1 1_2 \right> \left< 1_1 1_2 \right| \nonumber \\
\end{eqnarray}

\noindent then for energy eigenstates (\ref{eigenstates}), ${\vec{\bf \sigma}}_1 \cdot {\vec{\bf \sigma}}_2$ become still time independent:

\begin{eqnarray}
\left< u_1 \left| {\vec{\bf \sigma}}_1 \cdot {\vec{\bf \sigma}}_2 \right| u_1 \right> &=& 1 \nonumber \\
\left< u_2 \left| {\vec{\bf \sigma}}_1 \cdot {\vec{\bf \sigma}}_2 \right| u_2  \right> &=& 1 \nonumber \\
\left< u_3 \left| {\vec{\bf \sigma}}_1 \cdot {\vec{\bf \sigma}}_2 \right| u_3  \right> &=& 4j-1 \in (-1,1] \nonumber \\
\left< u_4 \left| {\vec{\bf \sigma}}_1 \cdot {\vec{\bf \sigma}}_2 \right| u_4  \right> &=& -4j-1  \in [-3,-1) \nonumber \\
\end{eqnarray}

\noindent noting  that the last eigenstate gives values corresponding to non separable states, so this quantity is a real entanglement witness.

\subsection{Entanglement and separability}

Using evolution operator we can verify that nevertheless  $\left| 0_1 0_2 \right>$, $\left| 1_1 1_2 \right>$ are invariant, the states:

\begin{eqnarray}
U(t') \left| 0_1 1_2 \right> &=&  e^{-i j t'} ((\cos t'-i b_- \sin t') \left| 0_1 1_2 \right>+ \nonumber \\ \nonumber
&& 2 i j  \sin t' \left| 1_1 0_2 \right>) \\ \nonumber
U(t') \left| 1_1 0_2 \right> &=&  e^{-i j t'} ((\cos t'+i b_- \sin t') \left| 1_1 0_2 \right>+ \\ \nonumber
&& 2 i j  \sin t' \left| 0_1 1_2 \right>) \\ 
\end{eqnarray}

\noindent have an interesting behavior. By calculate Schmidt coefficients, we note that these states are maximally entangled when: 

\begin{eqnarray}
\cos^2 t'+b_-^2  \sin^2 t'= 4j^2  \sin^2 t' \Rightarrow B_-^2=-4 J^2 \cos 2 R t \nonumber \\
\end{eqnarray}

\noindent this last, for non normalized time. It means:

\begin{eqnarray}
t_a &=& \frac{1}{2R} \arccos \left( -\frac{B_-^2}{4J^2} \right)+nT \nonumber \\
t_b &=& \frac{\pi}{R}-\frac{1}{2R} \arccos \left( -\frac{B_-^2}{4J^2} \right)+nT, n \in \mathbb{R} \nonumber \\
\end{eqnarray}

\noindent the period of the process is $T=\frac{2 \pi}{R}$ \footnote{$T=\pi/R$ is actually the period to get maximally entangled state, but they are not the same. The period to get the exactly the same state is $T=2 \pi/R$. Because of that, in one period it happens four times (except for $|B_- |=2J$).}. This happens only if  $\frac{B_-^2}{4J^2} \le 1$. Resulting states after of these times of evolution are respectively (dropping some unitary factors):

\begin{eqnarray}
\lefteqn {U(t_{a,b}) \left| 0_1 1_2 \right> =} \nonumber \\
&& \frac{1}{\sqrt{2}} ( e^{i \arctan(\frac{\sqrt{4J^2-B_-^2}}{B_-}{\rm sgn}(\tan R t_{a,b}))} \left| 0_1 1_2 \right>+
\left| 1_1 0_2 \right> ) \nonumber \\ 
\lefteqn{ U(t_{a,b}) \left| 1_1 0_2 \right> =} \nonumber \\
&& \frac{1}{\sqrt{2}} ( e^{-i \arctan(\frac{\sqrt{4J^2-B_-^2}}{B_-}{\rm sgn}(\tan R t_{a,b}))} \left| 1_1 0_2 \right>+
\left| 0_1 1_2 \right> ) \nonumber \\
\end{eqnarray}

\noindent where ${\rm sgn} (x)=\frac{|x|}{x}$ if $x \ne 0$. They reach two times maximal entanglement for each cycle, except if $\left|B_- \right|=2J$. In this case, the last both states become $\left|\beta_{01} \right>$, $\left|\beta_{10} \right>$ (just one of them depending on sign of $B_-$) at different times periodically. Inversely, this is the condition for the last Bell states become separable. 

Figure 2 shows the Von Neumann entropy for different initial states under this evolution (graphs are equal for $\left|0_1 1_2 \right>$ and for $\left|1_1 0_2 \right>$, nevertheless different states are reached). 

We don't show here evolution for initial maximally entangled states generated combining $\left| 0_1 1_2 \right>$ and $\left| 1_1 0_2 \right>$, but the results are similar. When $\left|B_- \right|=2J$ this state reaches with the same period a separable state, otherwise just reaches partial entanglement state. In the same way, they are invariant when $B_- \to 0$ or periodical when $B_- \to \infty$.

It´s important remark that the periodic behavior shown by last initial states it's only a partial view of phenomenon. Because of different energy eigenvalues arosen with the inhomogeneous magnetic field don't give rational quotients necessarily, this behavior for states initially entangled or separable have a non-periodic evolution in general. By example, taking as initial state ( $\theta$ parameter goes from  $0$ to $\frac{\pi}{2}$  , taking maximal entanglement in both, but having initial partial entanglement in intermediate values):

\begin{equation}
\left| \psi \right> = \sin \theta \left| \beta_{01} \right> - \cos \theta \left| \beta_{10} \right>
\end{equation}

\noindent and let it to have Ising interaction, we find that the Schmidt coefficients become after time $t'$:

\begin{eqnarray}
\lefteqn{\lambda_{1,2}=\frac{1}{2} ( 1 \pm (16j^2(1-4j^2) \sin^4 t' \sin^4 \theta +} \nonumber \\ 
&& \sin^2 2 \theta (\sin^2 2jt'+4j^2 \sin^2 t' \cos 4jt'-j \sin 4j t'))^{\frac{1}{2}} ) \nonumber \\
\end{eqnarray}

Here, $|b_-|= \sqrt{1-4j^2}$ has been used. Figure 3 show entanglement evolution of this more interesting state for different values of $\theta$  and  $j$, which exhibits the properties of this interaction. Appendix C shows some details about properties of entanglement evolution and conditions of periodicity of the separability for general cases.

By comparison with results of the before subsection, one can calculate using (\ref{evolspin}) for $\left| 0_1 1_2 \right>$, $\left| 1_1 0_2 \right>$:

\begin{eqnarray}
\left< 0_1 1_2 \right| {\vec{\bf \sigma}}_1 \cdot {\vec{\bf \sigma}}_2 (t') \left| 0_1 1_2 \right> &=& -(1+8jb_- \sin^2 t ) \in [-3,1]  \nonumber \\
\left< 1_1 0_2 \right| {\vec{\bf \sigma}}_1 \cdot {\vec{\bf \sigma}}_2 (t') \left| 1_1 0_2 \right> &=& -(1-8jb_- \sin^2 t ) \in [-3,1]  \nonumber \\
\end{eqnarray}

\noindent because  $8jb_- \in[-2,2]$. $\left< {\vec{\bf \sigma}}_1 \cdot {\vec{\bf \sigma}}_2 (t) \right> =-3$ when  $8jb_-=2$ in $t=\frac{2n+1}{2R}$,$n \in \mathbb{Z}$ for the first state and   $8jb_-=-2$ in $t=\frac{2n+1}{2R}$, $n \in \mathbb{Z}$ for the second state. These are precisely the times when these states become  $\left| \beta_{10} \right>$. In virtue of our previous result about entanglement witness, it tell us that this initially separable states reach an entangled stage by the action of the magnetic field . Precisely this is the reason because of one invariant state in the absence of magnetic field, now with inhomogeneous magneticfield we can achieve an separable form.

\section{Elemental procedures of control}
\subsection{Evolution loops}
Evolution loops were introduced by Mielnik \cite{mielnik1} and applied and extended in other directions of control by several authors \cite{ fernandez1, delgado1, delgado2} as simple operations to pursuit the specific behavior of quantum system. Note that here we don't introduce any stochastic element as considered in \cite{branczyk1, xi1}.

In some different sense we can reproduce this kind of effects by impose that the evolution operator after of some time $T$ of application become:

\begin{equation} \label{evoloop}
U(T)=e^{-i\phi} I
\end{equation}

\noindent where $I$ is the identity operator. Taking the evolution operator (\ref{evolop}), we note that the first condition that needs fulfill is $RT=n \pi$. In addition, analyzing the diagonal terms to fit to the form (\ref{evoloop}), we need impose the conditions:

\begin{eqnarray}
B_+ &=&\frac{2(s-m)J}{m+s-n}, \nonumber \\
B_- &=& \pm \frac{2J \sqrt{(2n-m-s)(m+s)}}{m+s-n},\nonumber \\
T &=& \frac{(m+s-n)\pi}{2J} 
\end{eqnarray}

\noindent with $n,m,s \in \mathbb{Z}$ and $0<n<m+s \le 2n$. We obtain:

\begin{equation}
U(T)=(-1)^n e^{-i JT} I
\end{equation}

It's clear that if  $p \in \mathbb{Z}^+$, then all cases with: $m,s,n \to pm,ps,pn$ are the same process but with $T \to pT$. Other cases are not physically equivalent. The fastest process, with $m+s-n=1$, appear as requiring stronger fields, nevertheless still is available without magnetic field by choosing $n,m,s=1$ : $B_+=B_-=0$,$T=\frac{\pi}{2J}$.  Figure 4 shows the entanglement evolution for the family of states:

\begin{equation}
\left| \varphi_\pm \right> = \sqrt{p} \left| 0_1 1_2 \right> \pm  \sqrt{1-p} \left| 1_1 0_2 \right>, p \in [0,1]
\end{equation}

\noindent under  these evolution loops.

\subsection{Transference of information}
Another control operation which could be induced is the transference or exchange of information between particles. Suppose that two particles in separate states begin to interact trough Ising interaction (\ref{hamiltonian}). It's possible that after some time $T$ this particles exchange their states? In order to get this effect, the evolution operator should be:

\begin{equation}
U(T)=e^{-i \phi} I_{1 \leftrightarrow 2}
\end{equation}

\noindent where $\phi$ is a phase and  $I_{1 \leftrightarrow 2}$ is the unitary exchange operator between particles 1 and 2:

\begin{eqnarray}
I_{1 \leftrightarrow 2}&=&  \left| 0_1 0_2 \right> \left< 0_1 0_2 \right|+\left| 0_1 1_2 \right> \left< 1_1 0_2 \right|+ \nonumber \\ 
&& \left| 1_1 0_2 \right> \left< 0_1 1_2 \right|+\left| 1_1 1_2 \right> \left< 1_1 1_2 \right|	\nonumber \\
\end{eqnarray}

This effect could, by example, induce the information transference:

\begin{eqnarray}
U(T)(\alpha \left|0_1 \right> + \beta \left| 1_1 \right>) \otimes \left| \psi \right>_2=e^{i\phi} \left|\psi \right>_1 \otimes (\alpha \left|0_2 \right> + \beta \left| 1_2 \right>) \nonumber \\
\end{eqnarray}

Fitting the Ising evolution operator to this last one, we obtain the conditions:

\begin{eqnarray}
B_+&=&2B_z=\frac{8Jm}{2n+1}, \nonumber \\
B_-&=& 0,\nonumber \\
T&=& \frac{(2n+1)\pi}{4J}
\end{eqnarray}

\noindent with $n \in \mathbb{Z}^+, m \in \mathbb{Z}$ so the evolution operator becomes:

\begin{eqnarray}
U(T)=i(-1)^n e^{-iJT}I_{1 \leftrightarrow 2}
\end{eqnarray}

It's remarkable that this phenomenon happens just with homogeneous field. In addition this effect could happen without magnetic field (choosing $m=0$), inclusively for the fastest process with $n=0$. This last it's only the fact that repeating two times this process one obtain of course an evolution loop, so this kind of operation is a subclass of Evolution loops but with one half of its period.

Figure 5 exhibit time evolution of entanglement for different initial separable states in one period of this effect depending on $p$ and $t$. It's remarkable that when states are less similar (more information to exchange), intermediate entanglement have stronger variations. Otherwise in the case of initial separable state the process require intermediate entangled states for the exchanging. All it means that while more information it's needed to exchange, systems require more intermediate entanglement increase.

\section{conclusions}
Study of the interaction between single pair of qubits is rich in complexity, specially when magnetic field is introduced. Erratical behavior is one of the principal aspects of this complexity because generalization to larger Ising strings is not easy, specifically that related with entanglement. Properties of the system exploited trough quantum control could let to emerge some useful aspects related not only with his driven but with transference of information, at least in the most direct approach (alternative or maybe related with presented by \cite{cai1}) because still the non-linear aspects of these strings could keep much more secrets and surprises. Study of control to stabilize this kind of systems and to drive it in order to transfer information in larger chains or rings could be considered trough of operations presented here.

\section*{Aknowledgements}
I gratefully acknowledge to Dr. Sergio Martinez-Casas about some fruitful discussions about use of Ising model in quantum cellular automatas which firstly inspired this study and to Dr. Bogdan Mielnik for comments about some basic quantum control operations, heritage from other areas of quantum control in our past works.

\section*{APPENDIX}
\subsection*{Appendix A: Symmetric Ising-Heisenberg model reliability}
A correct selection of parameters ${g_k}_{ij}$ shows that (\ref{ising}) it's always possible. By example selecting a symmetrical relative position of particles: $\vec{\mathbf{r}}=\frac{1}{\sqrt{3}}(1,1,1)$ and: 

\begin{eqnarray}
0=-g_{22} (g_{11}+g_{31})-g_{12} (g_{21}+g_{31})- \nonumber
(g_{11}+g_{21}) g_{32} \\ \nonumber
0=-g_{23} (g_{11}+g_{31})-g_{13} (g_{21}+g_{31})- 
(g_{11}+g_{21}) g_{33} \\ \nonumber
0=-g_{23} (g_{12}+g_{32})-g_{13} (g_{22}+g_{32})-
(g_{12}+g_{22}) g_{33} \\ \nonumber
g_{12}=-c, g_{13}=c, g_{23}=-c \\ \nonumber
g_{21}=c, g_{31}=-c, g_{32}=c \\  
\end{eqnarray}

\noindent with $c$ constant. If $c \in \mathbb{R}$ then $J>0$, and if $c \in \mathbb{I}$ then $J<0$.

\subsection*{Appendix B: Range for spected value of an observable for bipartite separable states}
The problem of evaluate the range of an observable for a bipartite separable state reduces to consider the function:

\begin{eqnarray}
f(x,y)&=&2(x \sqrt{1-y^2}+y \sqrt{1-x^2})^2+ \nonumber \\
&&\beta_1(2x^2-1)+\beta_2(2y^2-1)
\end{eqnarray}

\noindent with $x,y \in [0,1]$ and solving an optimization problem using calculus, which conduces to solutions for the extrema:

\begin{eqnarray}
\{ -\beta_+, 2-\beta_-,2+\beta_-,\beta_+ \}
\end{eqnarray}

\noindent where $\beta_+=\frac{\beta_1+\beta_2}{2}$ and $\beta_-=\frac{\beta_1-\beta_2}{2}$, when $x=0,1$ and/or $y=0,1$.

\subsection*{Appendix C: Periodicity of entanglement and separability in Ising model for two qubits} 

Taking a general bipartite state:

\begin{eqnarray}
\left| \varphi (0) \right>= \alpha \left| 0_1 0_2 \right> + \beta \left| 0_1 1_2 \right> +  \gamma \left| 1_1 0_2 \right> +  \delta \left| 1_1 1_2 \right>
\end{eqnarray}

\noindent with $|\alpha|^2+|\beta|^2+|\gamma|^2+|\delta|^2=1$. Representing this state in matrix form as:

\begin{eqnarray}
A(0)=\left(
\begin{array}{cc}
\alpha & \beta \\
\gamma & \delta
\end{array}
\right)
\end{eqnarray}

\noindent applying the Ising evolution we obtain for it:

\begin{eqnarray}
A(t)=\left(
\begin{array}{cc}
\alpha(t) & \beta(t) \\
\gamma(t) & \delta(t)
\end{array}
\right)
\end{eqnarray}

\noindent where:

\begin{eqnarray}
\lefteqn{ \alpha(t')= \alpha e^{-i(b_+-j)t'}} \nonumber \\
\lefteqn{ \beta(t')= e^{-ijt'} ( \beta (\cos t'-i b_- \sin t')+2ij \gamma \sin t')} \nonumber \\
\lefteqn{ \gamma(t')=} \nonumber \\
&& e^{-ijt'} ( \beta 2ij \sin t'+\gamma(\cos t'+i b_- \sin t')+\gamma 2ij \sin t') \nonumber \\
\lefteqn{ \delta(t')= \delta e^{i(b_++j)t'} }
\end{eqnarray}

\noindent the eigenvalues of:

\begin{eqnarray}
A(t)A^\dagger(t)=\left(
\begin{array}{cc}
a & b \\
c & d
\end{array}
\right)
\end{eqnarray}

\noindent will be the Schmidt coefficients:

\begin{eqnarray}
\lambda_{a,b}=\frac{1}{2} (1\pm \sqrt{1-4|\Delta(t')|^4})
\end{eqnarray}

\noindent with \footnote{By self $|\Delta(t) |$ is an entanglement measure because it's monotone between $0$ (separable) and $\frac{1}{\sqrt{2}}$ (maximally entangled state). Last means that if we write $S(|\Delta|)$, then $S$ is monotonically grown.}:

\begin{eqnarray}
| \Delta (t') |^2 = (ac-|b|^2)=|\alpha(t') \delta(t') -\beta(t') \gamma(t')|^2 
\end{eqnarray}

\noindent $\left| \varphi (t') \right>$ will be separable at time $t'$  iff  $\Delta(t')=0$. After some calculations we obtain:

\begin{eqnarray}
\lefteqn{ \Delta (t')= \alpha \delta e^{2ijt'}- \beta \gamma e^{-2ijt'}-} \nonumber \\
&& e^{-2ijt'}(( \beta^2+\gamma^2)ij \sin 2t' + \nonumber \\
&& (\beta^2-\gamma^2)2jb_- \sin^2 t' -8\beta \gamma j^2 \sin^2 t')
\end{eqnarray}

Using $\Delta \equiv \Delta(0)$, $\beta \equiv \gamma r e^{i \phi}$ one can write (if $\gamma \ne 0$) \footnote{Case $\beta=\gamma=0$ exhibit separability or entanglement invariance as was seen. Case $\gamma=0$ but $\beta \ne 0$ is similar to this by changing $r \to 1/r, \phi \to -\phi$. In addition, the study of right side of this equation can be restricted to $r \in (0,1)$ because cases with $r>1$ are obtained with the transformation $\phi \to - \phi, b_- \to - b_-$.}:

\begin{eqnarray}
\lefteqn{ F(r,\phi,j,t')= \frac {e^{-i\phi}}{\gamma^2} (\Delta (t')-\Delta e^{2ijt'})=} \nonumber \\
&& \frac {e^{-i(\phi-\phi_{\Delta'})}}{\gamma^2} (|\Delta (t')|-|\Delta| e^{2ijt'-i(\phi_{\Delta'}-\phi_{\Delta})})= \nonumber \\
&& -2r^2ij e^{i \phi -2ijt'} \sin t' (\cos t'-ib_-(j) \sin t') + \nonumber \\
&& 2r(i \sin 2jt'+4 e^{-2ijt'}j^2\sin^2t)- \nonumber \\
&& 2ije^{-i\phi-2ijt'}\sin t'(\cos t'+ib_-(j)\sin t') \nonumber \\
\end{eqnarray}

\noindent where $b_-(j)$ denotes the dependence of $b_-$ from $j$.

Here $\phi_\Delta$, $\phi_{\Delta'}$ are the phases of $\Delta, \Delta(t')$ respectively. This shows the non periodic behavior of separability and entanglement at least that $j \in \mathbb{Q}$. When the right side of this equation become zero for $t'>0$ and initial state is separable, then we have separability at time $t'$ again. But from the last formula, the same is true for any other value of  $|\Delta|$, because if $F(r,\phi,j,t')$ vanishes, it implies that the state (in general, different of initial state) at time $t'$ have the same value: $|\Delta(t') |=|\Delta|$. In addition their relative phase should be equal to $2jt'$ until some multiple of  $2 \pi$. Figure 6 show behavior of some cases of $F(r,\phi,j,t')$.

Some properties of $F(r,\phi,j,t')$ are remarkable. If $F(r,\phi,j,t')=F_r(r,\phi,j,t')+i F_i(r,\phi,j,t')$ are the real and imaginary parts of $F$, then for some $r$ and $\phi$ fix, we have that if  $F(j=\chi,t'=\tau')=0$ it means that until first order:

\begin{eqnarray}
F_r(\chi+{\rm d}j,\tau'+{\rm d}t')=\left. \frac{\partial F_r}{\partial j} \right| _{\chi,\tau'}{\rm d}j+\left. \frac{\partial F_r}{\partial t'} \right| _{\chi,\tau'}{\rm d}t' \nonumber \\
F_i(\chi+{\rm d}j,\tau'+{\rm d}t')=\left. \frac{\partial F_i}{\partial j} \right| _{\chi,\tau'}{\rm d}j+\left. \frac{\partial F_i}{\partial t'} \right| _{\chi,\tau'}{\rm d}t' \nonumber \\
\end{eqnarray}

If  $F_r(\chi +{\rm d}j ,\tau'+{\rm d}t')=F_i(\chi+{\rm d}j,\tau'+{\rm d}t')=0$ still, the condition of roots preservation, $\frac{{\rm d}t'}{{\rm d}j}$=0, gives:

\begin{eqnarray}
\left. \frac{\partial F_r}{\partial j} \right| _{\chi,\tau'} \left. \frac{\partial F_i}{\partial t'} \right| _{\chi,\tau'} -
\left. \frac{\partial F_r}{\partial t'} \right| _{\chi,\tau'} \left. \frac{ \partial F_i}{\partial j} \right| _{\chi,\tau'}
=0 \nonumber \\
\end{eqnarray}

In our case we obtain the result for last equation (taking $\tau ' =m \pi, \chi =\frac{n}{2m}$, the cases where we know that $F$ is zero for rational solutions):

\begin{eqnarray} \label{final}
-4 \pi nr(r^2-1) \sin \phi=0
\end{eqnarray}

\noindent it means that for cases $r=0,1$ or  $\phi=0, \pi$ the rational solutions $t'=m \pi,j=\frac{n}{2m},n,m \in \mathbb{Z}$ are preserved (not necessarily with the same values) for $j \in \mathbb{Q}'$ (see {\it d-f} in figure 6).

\section*{FIGURE CAPTIONS}
\subsection*{Figure 1}
Eigenvalues of energy depending on magnetic field ($B_-/J$ and $B_+/J$ parameters). Since $E_1/J$ (darkest) until $E_4/J$ (lightest). Note how $E_4/J$ is separated, suggesting different behavior which can be verified trough some properties.

\subsection*{Figure 2}
Entropy of entanglement behavior for the evolution of $\left| 0_1 1_2\right>$ and $\left| 1_1 0_2 \right>$ for different values of  $\left|B_-/J \right|$ in which can be reached maximally entangled states. Using $J=1$, periods shown are different: {\it a)} $\left|B_-/J \right|=0 \Rightarrow T=\pi$, {\it b)} $\left|B_-/J \right|=1 \Rightarrow T=\frac{2 \pi}{\sqrt 5}$, {\it c)} $\left|B_-/J \right|=2 \Rightarrow T=\frac{\pi}{\sqrt 2}$, {\it d)} $\left|B_-/J \right|=4 \Rightarrow T=\frac{\pi}{\sqrt 5}$. For higher values, as in {\it d)}, just partially entangled states are reached. Note in {\it c)}, the limit case, that just two times is reached a maximally entangled state, precisely $\left| \beta_{01} \right>$ or $\left| \beta_{10} \right>$. Note too that while magnetic field is more inhomogeneous the process is faster but far away of limit case $\left|B_-/J \right|=2$, the maximum of entropy reached goes to zero.

\subsection*{Figure 3}
Entanglement evolution for different values of $\theta$ and $j=\frac{J}{R} \in [0,\frac{1}{2}]$. Periodic behavior is only possible for rational $j$: {\it a)}$\frac{1}{16}$, {\it b)}$\frac{1}{8}$, {\it c)}$\frac{1}{4}$ and {\it d)}$\frac{3}{8}$.  Increasing $j$ we increase oscillations between values of entropy of entanglement, while for $\theta=\frac{\pi}{4}$ the maximum variation is reached in all cases. Evolution for $j$ irrational is shown in {\it e}), with $j=\frac{1}{\sqrt{7}} \approx \frac{3}{8}$, showing a non-periodic behavior.

\subsection*{Figure 4}
Time evolution of entanglement under evolution loops in one time period for different values of $p$. {\it a}) The case $m,n,s=1$ for $\left| \varphi_\pm \right>$; the case $n=3, m=0, s=2$ for {\it b)} $\left|\varphi_+ \right>$ and {\it c)} $\left|\varphi_- \right>$. Note in this last case the equivalence of inversion on the $p$ axis. In spite of that there are more than on cycle of periodicity of the graphs, the original states remain unchanged just when $t=T$.

\subsection*{Figure 5}
Time evolution of entanglement under exchange of information depending on $p$-parameter and time $t$, using $m,n=0$. For the initial state: {\it a)} $\left| \varphi_a \right> = (\sqrt{p} \left| 0_1 \right> + \sqrt{1-p} \left|1_1 \right>) \otimes \left|0_2 \right>$,  {\it b)} $\left| \varphi_b \right> = (\sqrt{p} \left| 0_1 \right> + \sqrt{1-p} \left|1_1 \right>) \otimes (\sqrt{p} \left| 0_2 \right> - \sqrt{1-p} \left|1_2 \right>)$,  {\it c)} $\left| \varphi_c \right> = (\sqrt{1-p} \left|0_1 \right> + \sqrt{p} \left| 1_1 \right>) \otimes (\sqrt{p} \left| 0_2 \right> - \sqrt{1-p} \left|1_2 \right>)$,  {\it d)} $\left| \varphi_d \right> = (\sqrt{1-p} \left|0_1 \right> + \sqrt{p} \left| 1_1 \right>) \otimes (\sqrt{p} \left| 0_2 \right> + \sqrt{1-p} \left|1_2 \right>)$. 
Graphs suggest that when more information should be transferred between parts, then more entangled become in the intermediate stage. Finally, just for comparison, the evolution for the entangled states:  {\it e)} $\left| \varphi_e \right> = \sqrt{p} \left|0_1 0_2 \right> + \sqrt{1-p} \left| 1_1 1_2 \right> $,  {\it f)} $\left| \varphi_f \right> = \sqrt{p} \left|0_1 1_2 \right> + \sqrt{1-p} \left| 1_1 0_2 \right> $. In the first case entropy of entanglement is unchanged, but in the second slightly variations appears when partial entanglement is weaker doubted to the same exchange of information-entanglement phenomena.

\subsection*{Figure 6}
Behavior of  $\left|F(r,\phi,j,t' )\right|$ for a set of values of $r$ (decreasing with darkness): {\it a-c)} shown for rational  $j =1/4$ exhibiting $2\pi$ periodicity of function and roots; note that  $\phi$ is related with symmetry between roots,  {\it d-f)} for irrational $j$, where periodicity is lost, it is remarkable that some roots appear accumulated near from $t'=4 \pi p$, $p \in \mathbb{Z}$ (specially for the cases $r=0,1$ or $\phi=0, \pi$ in accordance with result (\ref{final})). For rational $j$ it corresponds with $m=4$, which implies $j=\frac{n}{(2 \cdot 4)}$. Effectively, taking $n=3$ we note that  $j=\frac{3}{8}=0.375 \approx 0.377…= \frac{1}{\sqrt{7}}$. Anyway, repeatability of values for entanglement is normally present in Ising model, but they don't follow a periodic pattern.

\small  

\end{document}